\documentclass[prl,superscriptaddress,floatfix,twocolumn,showpacs,amsfonts,amsmath,amssymb,final]{revtex4}
\usepackage{graphicx} 
\usepackage{dcolumn} 
\usepackage{bm} 

\begin{document}

\title{Destruction of N\'eel order and local spin spirals in insulating La$_{2-x}$Sr$_x$CuO$_4$}

\author{Andreas L\"uscher}
\affiliation{School of Physics, University of New South Wales, Sydney 2052, Australia}
\author{Alexander I. Milstein}
\affiliation{Budker Institute of Nuclear Physics, 630090 Novosibirsk, Russia}
\author{Oleg P. Sushkov}
\affiliation{School of Physics, University of New South Wales, Sydney 2052, Australia}

\date{\today}

\begin{abstract}
Starting from the $t$-$J$ model, we derive an effective field theory describing the spin dynamics in the insulating phase of La$_{2-x}$Sr$_x$CuO$_4$, $x \lesssim 0.055$, at low temperature. Using Monte Carlo simulations, we show that the destruction of N\'eel order is driven by the single-hole localization length $\kappa$. A phase transition at $2\%$ doping is consistent with the value of $\kappa$ known from the variable range hopping conductivity. The static spin structure factor obtained in our 
calculations is in perfect agreement with neutron scattering data over the whole range of doping. We also demonstrate that topological defects (spin vortex-antivortex pairs) are an intrinsic property of the spin-glass ground state.
\end{abstract}


\maketitle

{\it Introduction.}
One of the most intriguing properties of La$_{2-x}$Sr$_x$CuO$_4$ (LSCO) is the static  incommensurate magnetic ordering at low temperature observed in elastic neutron scattering experiments. This ordering manifests itself as a scattering peak shifted with respect to the  antiferromagnetic position: ${\bf Q}={\bf Q}_{AF}+\delta{\bf Q}$, where ${\bf Q}_{AF}=\left(\pi,\pm\pi\right)$, setting the lattice spacing $a=1$ and using the tetragonal coordinate system. The incommensurate ordering is a generic feature of LSCO. According to experiments in the N\'eel phase, $x \lesssim 0.02$, the incommensurability is almost doping independent and directed along the orthorhombic $b$ axis, $\delta{\bf Q}\approx \pm 0.02 \sqrt{2}(\pi,-\pi)$, see Ref.~\onlinecite{matsuda02a}. In the spin-glass phase ($0.02 \lesssim x \lesssim 0.055$), the shift is again directed along the orthorhombic $b$ axis, but scales linearly with doping, $\delta{\bf Q}\approx \pm \sqrt{2} x(\pi,-\pi)$, see Refs.~\cite{wakimoto99a,matsuda00a,fujita02a}. Finally, in the underdoped  superconducting region ($0.055 \lesssim x \lesssim 0.12$), the shift scales linearly with doping and is directed along the crystal axes of the tetragonal lattice, $\delta{\bf Q}\approx 2x(\pm \pi,0)$ or $\delta{\bf Q}\approx 2x(0,\pm \pi)$, see Ref.~\onlinecite{yamada98a}.

These observations caused a renewal of theoretical interest in the idea of spin spirals in cuprates, both, from the phenomenological~\cite{hasselmann04a,juricic04a,lindgard05a,juricic06a} and the microscopic~\cite{sushkov04a,sushkov05a,luscher06a} point of view. The jump of the spiral direction at the insulator-superconductor transition ($x\approx 0.055$) was explained in Ref.~\onlinecite{sushkov05a} and a mechanism for the pinning of the spiral direction to the orthorhombic $b$ axis in the insulating phases (N\'eel and spin-glass regime) was suggested in Ref.~\onlinecite{luscher06a}. In this letter, we formulate an effective field theory that describes the spin-glass phase of LCSO at low temperature. We study the ground state, calculate the spin structure factor, and analyze the nature of the quantum phase transition from the N\'eel to the spin-glass state.

\begin{figure*}
\includegraphics[width=0.95\textwidth,clip]{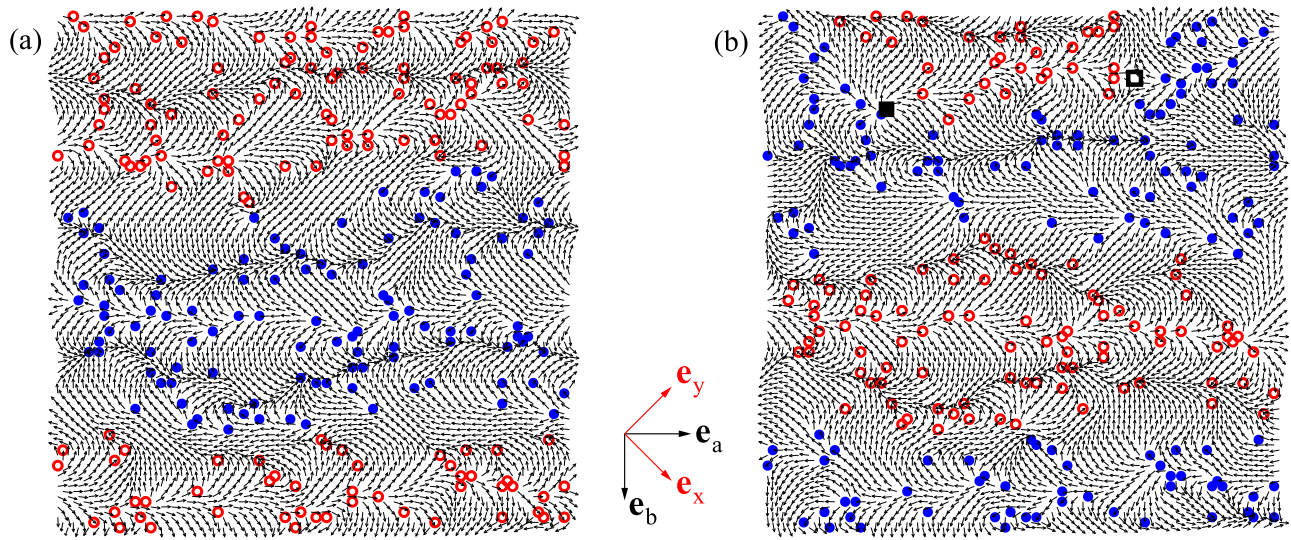}
\caption{\emph{(Color online). Characteristic ground state configuration of a particular realization at $x=0.05$. The system (a) without defects has higher energy than the same system (b) with a vortex-antivortex pair (squares). The impurity pseudospins $l_{i}$ are oriented along the $c$ axis. Full (open) circles correspond to values $-1$ ($+1$). Small arrows represent the ${\vec n}$-field. The system forms domains stretched along the $a$ direction, in which all the pseudospins are aligned in parallel.} \label{fig:groundstate}}
\end{figure*}

{\it Model.}
In insulating LSCO, holes are trapped by Sr ions and form hydrogen-like bound states $\psi\left({\bf r}\right) = \Psi \chi \left({\bf r}\right) = \Psi \sqrt{2/\pi} \kappa e^{-\kappa r}$,  where $\Psi$ is a two-component spinor describing the pseudospin~\cite{sushkov05a} and $\kappa \approx 0.4$ is the inverse localization length~\cite{chen95a}. We refer to these bound states as impurities~\cite{comment1}. The temperature corresponding to the binding energy is about $T_b \sim 100K$~\cite{sushkov05a}; hence at $T \ll T_b$ the charge degrees of freedom are frozen. Due to the small orthorhombic distortion of LSCO, impurities in the $S_{b}=\left(\pi/2,-\pi/2\right)$ hole pocket have lower energy than impurities in the vicinity of $S_{a}=\left(\pi/2,\pi/2\right)$. From the estimation of the pinning energy~\cite{luscher06a}, we conclude that for $T \ll T_{p} \sim 30K$, impurities are pinned to the $S_{b}$ valley and the spirals are directed along the $b$ axis, as explained in Ref.~\onlinecite{sushkov05a}. Since we consider $T \ll T_p$,  the relevant degrees of freedom are the spins of the Cu ions and the impurity pseudospins.

It is convenient to describe the staggered component of the copper spins by a continuous vector field ${\vec n}\left({\bf r}\right)$ within the framework of the non-linear $\sigma$-model (NLSM). Throughout this letter, we use the usual bold font for vectors in coordinate space and denote vectors acting in spin space by arrows. We use the same orthorhombic coordinate system in both  cases, i.e., ${\vec e}_\alpha={\bf e}_{\alpha}$, see Fig.~\ref{fig:groundstate}. Having in mind the above discussion of the relative degrees of freedom, we write the energy of a single layer of LSCO as
\begin{multline} \label{eq:en}
\frac{\rho_s}{2} \int d^2r 
\left\{\left[{\bf \nabla}{\vec n}\left({\bf r}\right)\right]^2+
\frac{D^2}{c^2} \left[n^{a}\left({\bf r}\right)\right]^2+\frac{\Gamma_{c}}{c^2} 
\left[n^{c}\left({\bf r}\right)\right]^2 \right\} \\
+\sqrt{2}g\sum_i\int d^2r \  \rho\left({\bf r-r_i}\right) \left[{\vec n}
\left({\bf r}\right) \times {\vec l}_i\; \right]\left({\bf e}_{b}\cdot{\bf \nabla}\right)
{\vec n}\left({\bf r}\right) \ ,
\end{multline}
where $n^\alpha$, $\alpha=a,b,c$, denotes the components of the ${\vec n}$-field, subject to the constraint ${\vec n}^2=1$. The first line in~(\ref{eq:en}) is the elastic energy which takes into account the
Dzyaloshinksi-Moriya (DM) and the XY anisotropies, with $D\approx 2$ meV and $\sqrt{\Gamma_c}\approx 4$ meV. We use $\rho_s \approx 0.18J$ for the spin stiffness and $c\approx 1.66J$ for the spin-wave velocity, with $J\approx 130$ meV. These parameters follow from neutron scattering data at zero doping~\cite{chovan00asilvaneto05a}. The second line in~(\ref{eq:en}) represents the interaction of the ${\vec n}$-field with the impurity pseudospins ${\vec l}_i=\Psi^\dag_i {\vec \sigma} \Psi_i$, see Ref.~\onlinecite{luscher06a}. The impurities are located at positions ${\bf r}_{i}$ and $\rho\left({\bf r}\right)=\chi^2\left({\bf r}\right)$. The coupling constant $g\approx J$ has been calculated previously within the extended $t$-$J$ model~\cite{sushkov04a}. Eq.~(\ref{eq:en}) describes the static limit of the effective field theory. This limit is sufficient to study ground state properties because~(\ref{eq:en}) results in a long-range interaction [see Eq.~(\ref{eq:ei}) below].

The anisotropies in Eq.~(\ref{eq:en}) pin the ${\vec n}$-field to a particular spatial direction. Because the XY term is larger than the DM anisotropy, $\sqrt{\Gamma_c} >D$, the staggered field ${\vec n}$ is coplanar and lies in the $ab$ plane. For a single CuO$_2$ layer at zero temperature, there is no mechanism that deflects ${\vec n}$ out of this plane. Substituting ${\vec n} = \left(n^{a}, n^{b}, n^{c}\right)=\left(\sin \theta, \cos \theta, 0\right)$ in Eq.~(\ref{eq:en}), the energy of the effective $O(2)$ NLSM reads
\begin{equation} \label{eq:e1}
\frac{\rho_s}{2} \int d^2r 
\left\{ \left[{\bf \nabla}\theta\left({\bf r}\right)\right]^2 + 2 {\cal M}\sum_i l_i \rho\left({\bf r}-{\bf r}_i\right) \left({\bf e}_{b} \cdot {\bf \nabla}\right) \theta\left({\bf r}\right) \right\},
\end{equation}
where $l_i=\Psi^\dag_i \sigma^c \Psi_i$ is an Ising variable taking the values $\pm 1$ and ${\cal M}=\sqrt{2} g/\rho_{s}\approx 8$. In~(\ref{eq:e1}) we have neglected the DM term $\frac{D^2}{c^2} \sin^2 \theta$. Despite its smallness, $D^2/c^2 \approx 10^{-4}$, this term is important in the N\'eel phase where it pins the ${\vec n}$-field to the $b$ axis, i.e., $\theta \approx 0$. However, in the spin-glass phase, the angle $\theta({\bf r})$ quickly varies at a scale much shorter than $l_{DM} =c/D \sim 100$ and since $\langle\sin^2 \theta\rangle=1/2$, the DM term can be safely neglected. Very importantly, Eq.~(\ref{eq:e1}) is exact in the sense that it is valid for arbitrary $\theta$ and not restricted to small angles. 

To integrate out the $\theta$-field, we start with a single impurity. The variation of Eq.~(\ref{eq:e1}) with respect to $\theta$ yields $-\nabla^2 \theta\left(\bf r\right) + {\cal M} l ({\bf e}_{b}\cdot{\bf \nabla})\rho\left(\bf r\right)  = 0$, which has the solution $\theta ({\bf r})=l \vartheta \left({\bf r}\right)$, where
\begin{equation} \label{eq:theta}
\vartheta \left({\bf r}\right) = 
\frac{{\cal M}}{2\pi}  \frac{{\bf e}_{b}\cdot{\bf r}}{r^2} 
\left\{\left(1+2 \kappa r\right) e^{-2 \kappa r}-1\right\} \ .
\end{equation}
Because of the linearity of the problem, the solution for $N$ impurities is the superposition 
\begin{equation} \label{eq:md}
\theta\left({\bf r}\right)=\sum_{i=1}^N l_i \vartheta\left({\bf r}-{\bf r}_{i}\right) \ . 
\end{equation}
Substituting this solution into Eq.~(\ref{eq:e1}) yields an expression of the energy in terms of the Ising pseudospins $l_i$ only
\begin{equation} \label{eq:ei}
E_{I} =\frac{\rho_s{\cal M}^2 \kappa^2}{2\pi} \sum_{i\neq j}^N l_i l_j \left\{F_{1}\left(\kappa r_{ij}\right) + \cos \left(2 \alpha_{ij}\right) F_{2}\left(\kappa r_{ij}\right)\right\} \ ,
\end{equation}
with $F_{1}\left(y\right) =-y^2 K_2\left(2 y\right)$ and $F_{2}\left(y\right) = 1/y^2-y K_3\left(2 y\right)-y^2 K_{2}\left(2 y\right)$. Here $\alpha_{ij}$ is the angle between the vectors ${\bf e}_{b}$ and ${\bf r}_{ij}={\bf r}_{i}-{\bf r}_{j}$, and $K_{n}$ are modified Bessel functions.  For $r \gg 1/\kappa$, the above expression is equivalent to the usual dipole-dipole interaction $\left[\propto \cos \left(2 \alpha_{ij}\right)/r_{ij}^2\right]$.

\begin{figure*}
\includegraphics[width=0.95\textwidth,clip]{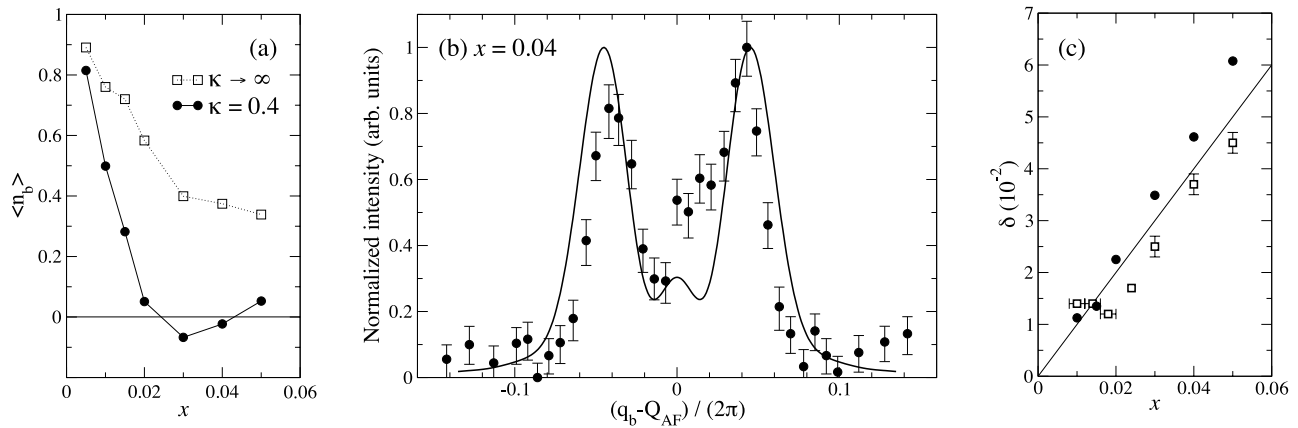}
\caption{\emph{(a) $\left\langle n^b \right\rangle$ as a function of doping $x$. For $\kappa=0.4$, N\'eel order is destroyed at $x_{c}\approx 0.02$, in agreement with experiments. For pointlike impurities ($\kappa \to \infty$), there is no phase transition at least up to $x=0.05$. (b) Neutron scattering probability $S_{\bf q}$ for $x=0.04$. Full circles correspond to experimental observations taken from Fig.~4 in Ref.~\onlinecite{fujita02a}, with normalized intensities. The curve represents our simulation, containing no fitting parameters. (c) Incommensurability $\delta$ (in reciprocal lattice units of the tetragonal lattice) as a function of doping. Our calculations (circles) are in good agreement with experimental measurements (squares) taken from Ref.~\onlinecite{wakimoto00a,matsuda00a,matsuda02a}, see Fig.~6 of Ref.~\onlinecite{matsuda02a}. Theoretical points for the N\'eel phase are taken from Ref.~\onlinecite{luscher06a}} 
\label{fig:neutron}}
\end{figure*}

{\it Ground state and destruction of N\'eel order}.
We perform classical Monte Carlo simulation to find the ground state of the Ising pseudospins described by Eq.~(\ref{eq:ei}) using the same algorithm as in Re.~\onlinecite{luscher06a}. Because samples of LSCO consist of many CuO$_{2}$ layers with randomly distributed dopants, we average observables over many realizations of random impurity distributions. For a particular realization, we consider up to $N=200$ Ising pseudospins $l_{i}$ on a square lattice of size $L=\sqrt{N/x}$. In order  to minimize finite-size effects, we orient the lattice along the orthorhombic coordinate system, apply periodic boundary conditions along the $a$ axis and extract relevant quantities only from the central quarter of the system. The $b$ direction is left open, in order not to impose an artificial constraint on the spiral pitch. Once the Ising ground state is found, the ${\vec n}$-field is determined according to Eq.~(\ref{eq:md}). Note that $\theta({\bf r})$ in~(\ref{eq:md}) can always be shifted by a constant $\theta_{0}$. We set $\theta_{0}=0$ and therefore have $\theta=0$ and ${\vec n}={\vec e}_{b}$ in the undoped system. Fig.~\ref{fig:groundstate}(a) shows a characteristic spin-glass ground state for a particular realization at doping $x=0.05$. Open (full) circles show the Ising pseudospins directed parallel (antiparallel) to the $c$ axis (orthogonal to the plane). The ${\vec n}$-field is represented by arrows. The systems forms large domains, stretched along the $a$ axis, in which the pseudospins all point in the same direction. Fig.~\ref{fig:groundstate}(a) nicely illustrates the concept of spiral formation. Despite the random distribution of impurities, one easily recognizes horizontal lines along which the ${\vec n}$-field remains almost constant. In contrast to a uniform spiral, these lines are not exactly parallel to the $a$ axis, but meander around the impurities.

The presence of N\'eel order can be identified as a non-zero $\left\langle n^b \right\rangle$. For  a given doping, the average is taken over the central quarter of the system and over many realizations of random impurity positions. We first performed calculations for $\kappa \to \infty$ (pointlike impurities) and found that N\'eel order is not destroyed at least up to $x=0.05$, see Fig.~\ref{fig:neutron}(a).
In this case, the pseudospins align in an antiparallel pattern along the $b$ direction, on a scale of the order of the separation between impurities. The angle $\theta$ is thus small and the resulting $n$-field never completes a full rotation. For finite $\kappa$, the situation is qualitatively different because for sufficiently small distances between impurities, the combination in brackets $\{...\}$  in Eq.~(\ref{eq:ei}) is always negative. This favors parallel alignment of the pseudospins and hence leads to the formation of parallel Ising domains, see Fig.~\ref{fig:groundstate}(a). The domains have a finite width because the long-range tail of~(\ref{eq:ei}) still favors antiparallel alignment along the $b$ direction. The width $w$ depends on doping and on $\kappa$, but also on the size of the lattice as $w\propto \sqrt{L}$. The domains thus become macroscopic in the thermodynamic limit. The mechanism for the domain formation is exactly the same as for ferromagnets~\cite{kittel95a}. We will show in the following paragraph that topological defects lead to finite, but still very large domain sizes. The doping dependence of $\left\langle n^b \right\rangle$ for $\kappa=0.4$ is shown in Fig.~\ref{fig:neutron}(a). The phase transition takes place at  $x_c \approx 0.02$, in very good agreement with experiments. The small deviations from zero at $x \ge 0.02$ in Fig.~\ref{fig:neutron}(a) are due to finite size effects. It is remarkable that the correct critical concentration $x_{c}$ is obtained for $\kappa \approx 0.4$ that is known from the variable range hopping conductivity~\cite{chen95a}. 

{\it Topological defects}.
There are two kinds of topological defects in a 2D $O(3)$ NLSM, vortices and instantons~\cite{polyakov87a}. Instantons lead to non-coplanar spin configurations that, according to our calculations, increase the energy because the system prefers coplanar pseudospin arrangements. Instantons are therefore not present in the ground state. For vortices, the situation is different. Let us consider solutions of the Laplace equation $\nabla^2 \theta=0$ of the form $ \theta_v\left({\bf r}\right)=\sum_{j=1}^M Q_{j} \arg \left[x-X_{j}+i \left(y-Y_{j}\right)\right]$. Here ${\bf R}_{j}=\left(X_{j},Y_{j}\right)$ and $Q_{j} \in\mathbb{Z}$ are the positions of the impurities and their topological charges (winding numbers), respectively. The  energy is minimal for vanishing total topological charge with $Q_j=\pm 1$. In this case, we have $M/2$ pairs of vortices with opposite winding numbers. For large average separation between vortices, $R \gg 1$, the associated energy calculated with logarithmic accuracy is~\cite{polyakov87a}
\begin{equation} \label{eq:ev}
E_{V} = M \pi \rho_{s} \ln R \ .
\end{equation}
Because this energy is positive, vortices never appear in the ground state of the NLSM without impurities. However, in a doped system, one also has to take into account the interaction energy $E_{IV}$ between impurities and vortices. From Eq. (\ref{eq:e1}), we find
\begin{equation} \label{eq:eiv}
E_{IV}= \rho_{s} {\cal M} \sum_{i,j=1}^{N,M} l_{i} \ Q_{j} 
\frac{{\bf e}_{a} \cdot {\bm \xi}_{ij}}{\xi_{ij}^2} 
\left[1-e^{-2 \kappa \xi_{ij}}\left(1+2 \kappa \xi_{ij}\right)\right] \ ,
\end{equation}
where ${\bm \xi}_{ij}=\left({\bf R_{j}}-{\bf r_{i}}\right)$ are the positions of the vortices with respect to the impurities. The interaction energy~(\ref{eq:eiv}) is large and negative because of the domains formed in the spin-glass phase, in which all pseudospins point in the same direction. It therefore favors the creation of vortices at zero temperature. The total energy is now given by the sum of~(\ref{eq:ei}), (\ref{eq:ev}) and (\ref{eq:eiv}). An illustration of a system with one vortex-antivortex pair is shown in Fig.~\ref{fig:groundstate}(b). Compared to the same realization without defects [Fig.~\ref{fig:groundstate}(a)], the energy is around $10\%$ lower. The  presence of vortices renders the problem of finding the ground state quite difficult. In addition to minimizing the energy of the Ising pseudospins, one also has to check if vortex-antivortex pairs can lower the energy, and if so, optimize their positions. Nevertheless, it is possible to estimate the optimal distance between defects in the thermodynamic limit. Because of the long-range nature of the interaction~(\ref{eq:eiv}), the energy gain is proportional to $R$. Let us denote the energy of the system without vortices by $E=E_{I}$ and use $\tilde E = \tilde E_{I} + E_{IV} +E_{V} $ for a system with $M$ defects. The energy gain due to vortices can then be expressed as  $\Delta E =  \tilde E-E = -\gamma x M R + M \pi \rho_{s} \ln R$. By comparing the energies of systems with $4$ and $16$ vortices to the same realizations without defects, we find that this formula gives the correct $R$ dependence. The parameter $\gamma$ is weakly doping dependent, at $x=0.02$, we obtain $\gamma \approx 2.0$, leading to an optimal distance $R_\text{opt} \approx 130$ in the thermodynamic limit. At $x=0.05$, we find $\gamma \approx1.1$ from which we deduce $R_\text{opt} \approx 90$. We thus conclude that vortex-antivortex pairs break up the domains and lead to a parquet-like arrangement of Ising pseudospins with a natural domain size $R\approx100$.

{\it Static spin structure factor and neutron scattering.}
Let us finally calculate the structure factor 
\begin{equation}
S^{\alpha\beta}_{\bf q}= \frac{1}{L^2} \sum_{i,j} e^{i {\bf q} \cdot \left( {\bf r}_{i}-{\bf r}_{j}\right)} n^\alpha\left({\bf r}_{i}\right) n^\beta\left({\bf r}_{j}\right) \ ,
\end{equation}
as well as the sum $S_{\bf q}=\delta_{\alpha\beta}S^{\alpha\beta}_{\bf q}$, which is related to the neutron scattering cross section. Fig.~\ref{fig:neutron}(b) shows experimental data taken by Fujita~{\it et al.}~\cite{fujita02a} on a sample with doping $x=0.04$ together with our theoretical results for $S_{\bf q}$. The agreement between theory and experiment is quite remarkable, especially given the fact that it is not restricted to a particular sample, but can be observed over a broad range of doping. This latter finding is illustrated in Fig.~\ref{fig:neutron}(c), which shows the incommensurability $\delta$, defined as half the distance between the peaks, as a function of doping. Expressed in reciprocal lattice units of the tetragonal structure, the incommensurability is in good approximation proportional to doping, $\delta \approx x$. Because of their low density, topological defects do not have a major influence on $S_{\bf q}$, and only lead to some broadening of the peaks.

To conclude, we have developed a low energy effective field theory that describes the magnetic structure of La$_{2-x}$Sr$_x$CuO$_4$ in the spin-glass phase ($x \lesssim 0.055$). We have shown that the staggered component of the copper spins is confined to the $ab$ plane, due to the XY anisotropy. The static spin structure factors obtained in our approach are in excellent agreement with neutron scattering data, over a whole range of doping. We have analyzed the nature of the quantum phase transition from the N\'eel to the spin-glass phase and have also shown that topological defects (spin vortices) play a significant role in the ground state of the spin-glass phase. 

\acknowledgments{We would like to thank A.~N. Lavrov for valuable discussions. A.~I.~M. gratefully acknowledges the School of Physics at the University of New South Wales for warm hospitality and financial support during his visit. The Monte Carlo simulations have been performed on computers belonging to the Institut Romand de Recherche Num\'erique en Physique des Mat\'eriaux (IRRMA) in Switzerland.}


\begin{thebibliography}{17}
\bibitem{comment1} We would like to emphasize that impurities are an intrinsic part of the system.
\end{thebibliography}
\end{document}